\documentclass[twocolumn,showpacs,preprintnumbers,amsmath,amssymb]{revtex4}

\usepackage{graphicx}
\usepackage{dcolumn}
\usepackage{bm}
\usepackage{epsfig}
\usepackage{CJK}
\usepackage{bm}

\begin{document}


\title{Scaling Behaviors and Novel Creep Motion of Flux Lines under AC Driving}


\author{Wei-Ping Cao$^{\dagger,\ddagger}$, Meng-Bo Luo$^{\dagger,\ddagger}$
and Xiao Hu$^{\dagger}$}

\affiliation{ $^{\dagger}$WPI Center for Materials
Nanoarchitectonics, National Institute for Materials Science,
Tsukuba 305-0044, Japan
\\ $^{\ddagger}$Department of Physics, Zhejiang University, Hangzhou 310027,
China}

\date{\today}

\begin{abstract}
We performed Langevin dynamics simulations for the \textit{ac}
driven flux lines in a type II superconductor with random point-like
pinning centers. Scaling properties of flux-line velocity with
respect to instantaneous driving force of small frequency and around
the critical \textit{dc} depinning force are revealed successfully,
which provides precise estimates on dynamic critical exponents. From
the scaling function we derive a creep law associated with the
activation by the regular shaking. The effective energy barrier
vanishes at the critical dc depinning point in a square-root way
when the instantaneous driving force increases. The frequency plays
a similar role of temperature in conventional creep motions, but in
a nontrivial way governed by the critical exponents. We have also
performed systematic finite-size scaling analysis for flux-line
velocity in transient processes with \textit{dc} driving, which
provide estimates on critical exponents in good agreement with those
derived with ac driving. The scaling law is checked successfully.
\end{abstract}


 \maketitle

\noindent \textit{Introduction. --} Dynamics of elastic manifolds
driven through a random medium is one of the rich paradigms in
condensed matter physics. Important examples range from
ferromagnetic or ferroelectric domains, charge-density waves,
frontiers of fluids in porous media, Wigner crystal, and flux lines
in type II superconductors \cite{Kardar,Fisher_PR,Nattermann_AP}.
The notion that the depinning transition by a dc driving at zero
temperature in these systems exhibits properties similar to the
critical phenomena of a second-order phase transition at thermal
equilibrium has been verified successfully in many systems
in this class \cite{Fisher}, while a possibility was raised
recently on the absence of divergent correlation
length below the depinning point \cite{Giamarchi}.

Special attentions have been paid to the dynamic phenomena of flux
lines in bulk superconductors with randomly
distributed defects \cite{Blatter_RMP}. It is because, first of all,
the system is important for applications of superconductivity, such
as transport of large current and generation of strong magnetic field. Flux
lines are also unique with respect to the depinning transition due
to the two-dimensional (2D) dynamic degrees of freedom under a driving transverse to
the direction of magnetic field, which hampers an analytic treatment
based on the functional renormalization group \cite{FRG}. Computer
simulations based on Langevin dynamics
\cite{Brandt,Ryu,Reichhardt,Zimanyi,Olive,Dominguez,LuoHu} were
performed in order to fill the gap.

While the critical depinning properties of flux lines under \textit{dc} driving
were addressed quite satisfactorily (see for example \cite{LuoHu}), behaviors
under \textit{ac} driving have not yet been fully
explored so far, despite of its importance and convenience as an
experimental probe. Fortunately, a scaling theory for small
frequencies around the critical dc depinning point has been
formulated based on analysis on an elastic string embedded in 2D
random medium \cite{Nattermann}, which provides a facet into this
problem and motives our present study.

We performed computer simulations for the ac driven flux lines in
presence of quenched random point-like pins based on the
Langevin dynamics. Scaling properties of flux-line velocity with
respect to instantaneous driving force of small frequency and around
the critical dc depinning force are revealed successfully based on
the scaling theory \cite{Nattermann}, which provides precise
estimates on dynamic critical exponents. From the scaling function
we found a creep law associated with the activation by the regular
shaking, with the energy barrier vanishing at the critical dc
depinning point in a square-root way when the instantaneous driving
force increases. The frequency appears in the creep law, instead of
temperature in a similar creep motion under dc driving
\cite{Blatter,LuoHu}, in a nontrivial way governed by the critical
exponents. In order to check possible effects from finite systems in
simulations on accuracy of critical exponents, we also simulated two
transient dynamic processes under \textit{dc} driving and performed
systematic finite-size scaling analysis. It is found that the
estimates on the critical exponents converge nicely, and moreover
satisfy the scaling law \cite{LHTang}.

\vspace{3mm}

\noindent \textit{Model and simulation technique. --}
The model system is a stack of superconducting layers with a perpendicular magnetic field,
same as that used in Ref.~\cite{LuoHu}.  The overdamped equation of motion of the
\(i\)th vortex at position ${\bf r}_i$ is

\vspace{-3mm}
\begin{equation}
   \eta \dot{{\bf r}}_i = -\sum\limits_{j\neq i} \nabla_i
   U^{\mathrm {VV}}({\bf r}_{ij})- \sum\limits_{p} \nabla_i
   U^{\mathrm{VP}}({\bf r}_{ip}) + {\bf F} + {\bf F}_{\mathrm {th}},
\label{Ldynamics}
\end{equation}

\noindent with $\eta$ the viscosity coefficient. The intraplane
vortex repulsion is given by the modified Bessel function, the
interplane vortex attraction is approximated by a spring potential
between two vortices belonging to the same flux line and sitting on
adjacent planes, and pinning centers are modeled by Gaussian
potentials with typical size $R_{\mathrm p}$ distributed at random
positions but with constant pinning range and dimensionless pinning
strength $\alpha$; ${\bf F}$ is the Lorentz force uniform over the
system but varies with time as $F(t)=A\sin\omega t$, and ${\bf
F}_{\mathrm{th}}$ is the thermal noise force. In this work, the
units for length, energy, temperature, force, time, and velocity are
taken as $\lambda_{ab}$, $d\epsilon_0$, $d\epsilon_0/k_{\mathrm B}$,
$d\epsilon_0/\lambda_{ab}$, $\eta\lambda_{ab}^2/d\epsilon_0$($\equiv
\tau_0$), and $d\epsilon_0/\eta \lambda_{ab}$, respectively, where
$\epsilon_0 = \phi_0^2/2\pi \mu_0 \lambda_{ab}^2$, $\lambda_{ab}$
the magnetic penetration depth, and $d$ the thickness of the
superconducting layer. The results will be given in dimensionless
units hereafter. The dynamic equation (\ref{Ldynamics}) is
integrated with the 2nd order Rugen-Kutta algorithm with \(\Delta
t=0.01\sim 0.02\). More details about our model system can be found
in Ref.~\cite{LuoHu}.

In simulations there are totally $N_{\mathrm v} = 180$ flux lines in
the system of lateral size $L\times L=30 \times 30$ and 20 layers in
the $c$ axis, with periodic boundary conditions. Intra-plane
vortex-vortex repulsions are cut off at $r_{\mathrm{cut}}=6$. Each
layer contains $N_{\mathrm p} = 900$ randomly distributed point pins
with $R_{\mathrm p} = 0.22$ and $\alpha=0.2$, associated with a
vortex glass at thermal equilibrium \cite{LuoHu}. For different system sizes, the
densities of the flux lines and the point pins are fixed. In the
present work we concentrate on zero temperature.

\vspace{3mm}

\noindent \textit{ac driving. --} We first investigate motions of
flux lines under an ac driving. Double hysteresis loops
\cite{Nattermann} are observed as displayed in the inset of
Fig.~1(a) for $\omega=0.01\pi$. It is clear that for a finite
$\omega$ the velocity of the system is nonzero even when the
instantaneous driving force is below the critical dc depinning force
$F_{\rm c0}$ derived from motions of flux lines under \textit{dc}
driving \cite{LuoHu}. We therefore refer to $F_{\rm c0}$ simply as
critical depinning force hereafter.

As the frequency $\omega$ approaches zero, the double loops shrink
to the depinning curve of the dc driving \cite{LuoHu}, as seen in
the main panel of Fig.~1(a) where the instantaneous force dependence
of the velocity is shown around $F_{\rm c0}$ in the increasing
branch. According to the scaling theory \cite{Nattermann}, the
behavior of the system for $\omega<\omega_p=F_{\rm c0}/\eta h\simeq
0.08\pi$ is governed by the critical properties of the depinning
fixed point

\vspace{-3mm}
\begin{equation}
v(t)=\omega^{\beta/\nu z} \phi\left[(F(t)/F_{\rm
c0}-1)\omega^{-1/\nu z}\right]; \label{wscaling}
\end{equation}

\noindent here the critical exponents are defined by the onset of
the velocity $v\sim (F/F_{\rm c0}-1)^{\beta}$ under dc driving, the
divergent correlation length $\xi\sim |F/F_{\rm c0}-1|^{-\nu}$, and
the growth of correlation length with time $\xi\sim t^{1/z}$ at
$F=F_{\rm c0}$. It is requested that $\phi(x)\sim x^{\beta}$ for
$x\rightarrow \infty$ in order to recover the steady depinning
behavior.

\begin{figure}[t]
\psfig{figure=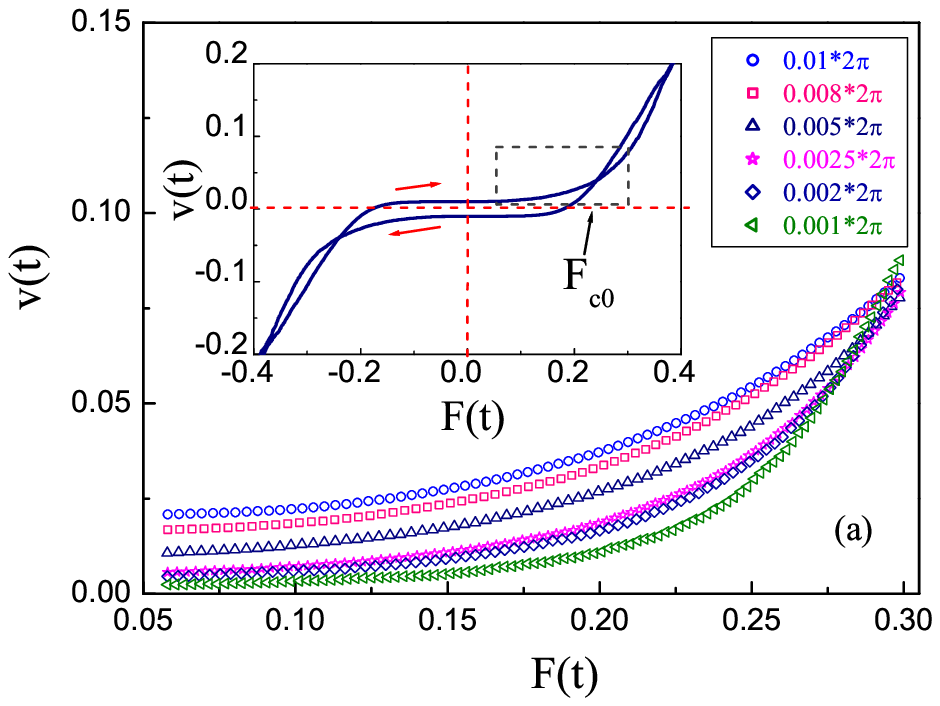,width=8cm}
\psfig{figure=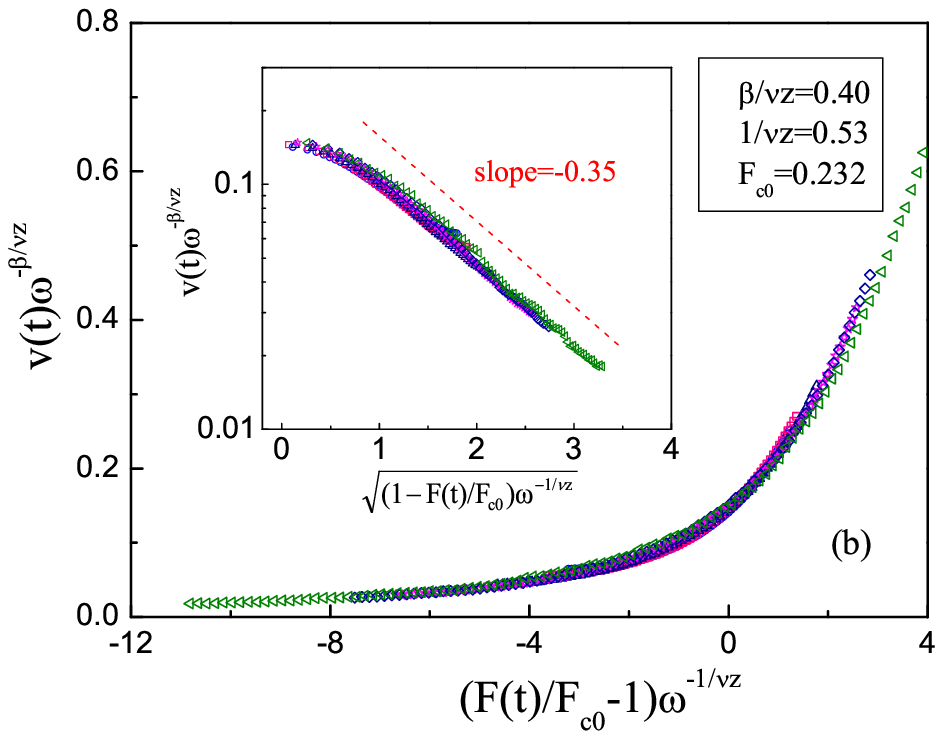,width=8cm} \caption{(color online). (a)
Hysteretic loops of the velocity of flux lines under ac driving
$F(t)=0.4\sin(0.01\pi t)$ (inset) and the part around $F_{\rm c0}$
in the increasing branch (dashed rectangular in the inset) for
several frequencies (main panel). (b) Scaling plot according to
Eq.(\ref{wscaling}) (main panel) and asymptote of the scaling
function $\phi(x)$ in Eq.(\ref{wscaling}) as $x\rightarrow -\infty$
(inset). Simulations were performed with $L=30$ and numbers of
samples are 80-200 depending on the frequency.}
\end{figure}

The scaling plot according to Eq.(\ref{wscaling}) sees a good
collapsing with $\beta/\nu z=0.40\pm 0.01$ and $1/\nu z=0.53\pm
0.02$, and $F_{\rm c0}=0.232\pm 0.002$ as displayed in the main
panel of Fig.~1(b). The estimates on the exponent $\beta=0.76\pm
0.02$ and $F_{\rm c0}$ agree well with the previous results based on
dc driving at low but finite temperatures ($\beta=0.75\pm 0.01$ and
$F_{\rm c0}=0.232\pm 0.001$) \cite{LuoHu}.

A clear asymptotic behavior is observed for the scaling function
$\phi(x)$ as $x\rightarrow -\infty$ in the main panel of Fig.~1(b).
As replotted in the inset of Fig.~1(b),  the
asymptote is well described by

\vspace{-3mm}
\begin{equation}
v(t)\sim \omega^{\beta/\nu z} \exp{\left[-\frac{0.35\sqrt{1-F(t)/F_{\rm
c0}}}{\omega^{1/2\nu z}}\right]}, \label{creep}
\end{equation}

\noindent for $F(t)<F_{\rm c0}$. The motion of flux lines is creep
like, with an effective energy barrier $0.35\sqrt{1-F(t)/F_{\rm c0}}$ which vanishes
 when the instantaneous driving force approaches $F_{\rm c0}$
from below. The creep motions of flux lines here are caused by the
regular shaking of the ac driving, instead of the thermal
activations at finite temperature \cite{Blatter,LuoHu}. Because of
the random pinning potential landscape, a regular shaking
contributes to activation of flux lines over pinning barriers in a
complex way governed by the critical properties, both steady and
dynamic, of the system, which is captured by the frequency
dependence in the creep law (\ref{creep}). A theory on the
square-root suppression of the energy barrier is not available at
this moment.

To end this part we notice that, as seen in the scaling theory (\ref{wscaling}),
the critical exponents
$\nu$ and $z$ appear in the form of product in the dynamics under ac
driving, which leaves a full description of the critical dynamics unavailable.
On the other hand, because of the hysteretic response of flux lines to ac driving, it is still
hard to make any serious analysis on finite-size effects on the
accuracy of the critical exponents with the computing resources
available at this moment. These issues are addressed in what follows.

\vspace{3mm}

\begin{figure}[t]
\psfig{figure=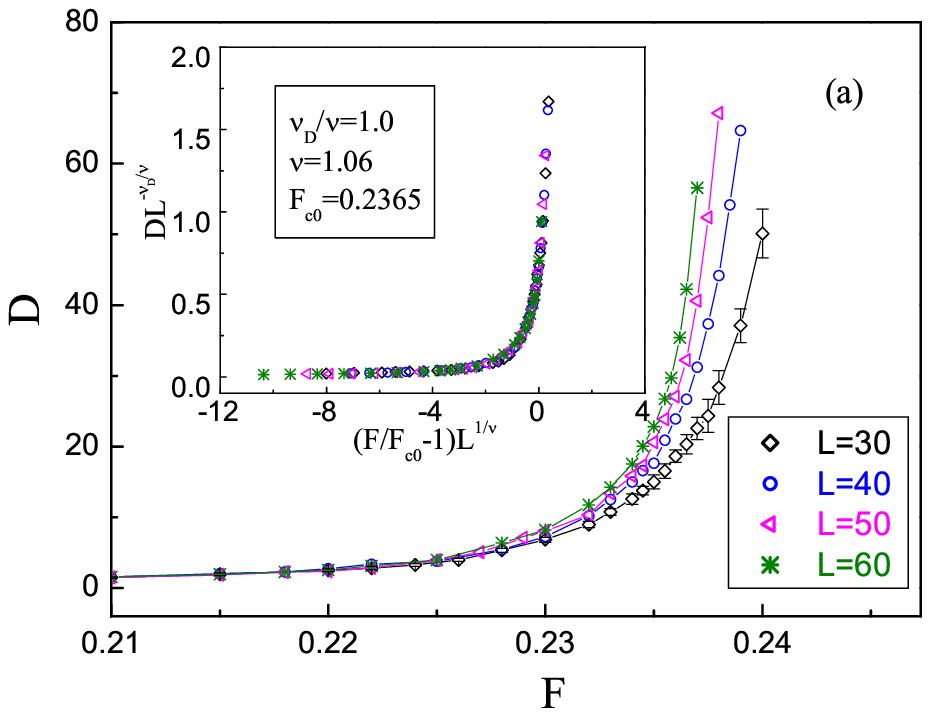,width=8cm}
\psfig{figure=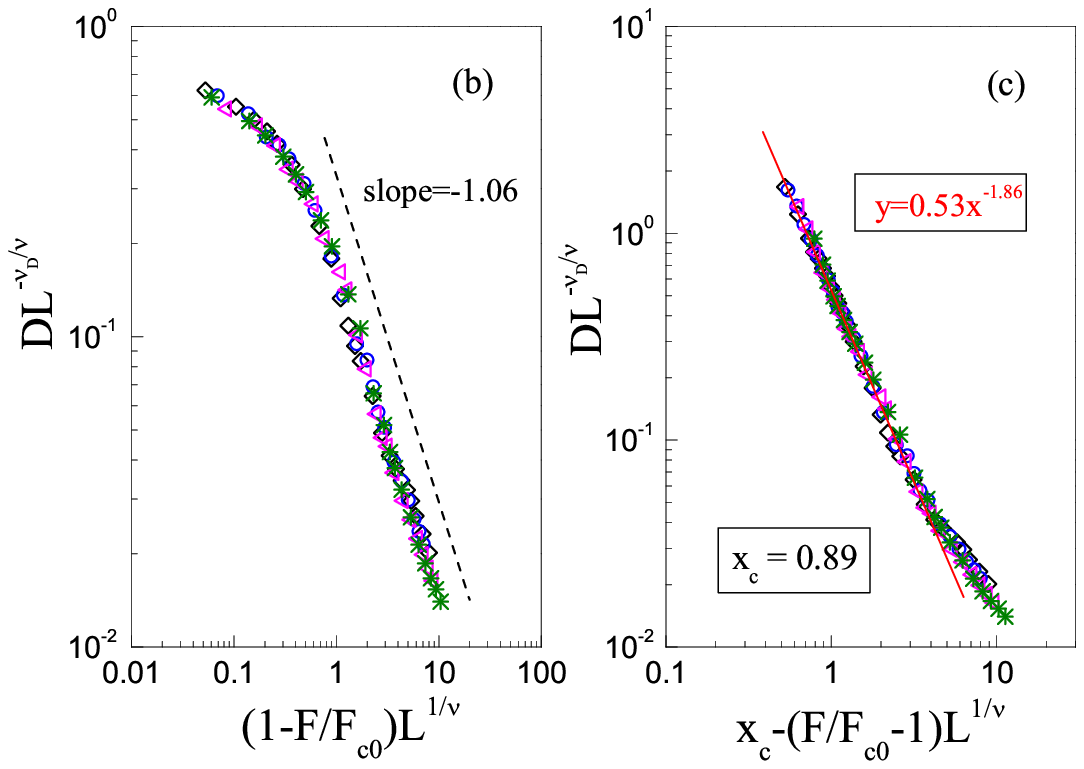,width=8cm} \caption{(color online). (a)
Traveling distance of the flux lines after relaxing from a perfect
triangular lattice for several finite systems (main panel) and
scaling plot based on the finite-size scaling ansatz
(\ref{Dscaling}) (inset). (b) Asymptote of the scaling function
$S(x)$ in the scaling theory (\ref{Dscaling}) for $x\rightarrow
-\infty$ and (c) for $x\rightarrow x_{\rm c}$. Numbers of samples
are 10-50 depending on the system size. Error bars are given for the
system of $L=30$ and those for other system sizes are similar.}
\end{figure}

\noindent \textit{Onset of collective pinning. --}
Let us now investigate a process associated with the onset of
collective pinning, which involves only the exponent $\nu$.
We lay at $t=0$ a perfect triangular lattice of flux lines on the
random medium, and start to drive the flux
lines by a \textit{dc} force. The flux-line lattice deforms during
traveling when the individual flux lines adapt to the random potential
landscape. As the result, the dragging caused by random pinnings
becomes stronger due to the collective pinning mechanism \cite{LO}.
When the dc driving force
is below the critical depinning one, the flux lines should be
stopped as a whole after traveling a certain distance. This
traveling distance depends on how far the driving force is from the
critical point, and is determined by the critical properties
of the depinning fixed point.

We have simulated this process in several finite systems, and the
results are displayed in the main panel of Fig.~2(a). It is clear
that, for a given finite system, the traveling distance diverges
(since a periodic boundary condition is adopted) as the driving
force approaches a critical value given by the system size. The
critical force decreases as the system size increases. The
critical behavior for an infinite system is given by

\begin{equation}
D\sim (1-F/F_{\rm c0})^{-\nu_D},
\label{D}
\end{equation}

\noindent with an exponent $\nu_D$, and the convergence from
finite systems to the infinite system should be described by
the finite-size scaling theory \cite{Roters}:

\vspace{-3mm}
\begin{equation}
D=L^{\nu_D/\nu} S\left[(F/F_{\rm c0}-1)L^{1/\nu}\right].
\label{Dscaling}
\end{equation}

\noindent The scaling plot according to Eq.(\ref{Dscaling}) is shown
in the inset of Fig.~2(a). From the successful scaling plot, we
obtain $\nu_D/\nu=1.00\pm 0.05$, $\nu=1.06\pm 0.04$, and $F_{\rm
c0}=0.2365\pm 0.0005$. The present estimate on the critical
depinning force agrees with the above one derived for ac driving and
that in the previous work for dc driving \cite{LuoHu}, but with the
highest precision since finite-size effects are taken into account
here.

Two features of the scaling function $S(x)$ are observed as follows:
(i) As shown in Fig.~2(b), $S(x)\sim (-x)^{-\nu_D}$ for
$x\rightarrow -\infty$ as requested by the scaling theory, which
achieves the critical behavior (\ref{D}). (ii) As shown in
Fig.~2(c), $S(x)\simeq a(x_{\rm c}-x)^{-p}$ for $x\rightarrow x_{\rm
c}$, with $a\simeq 0.53$, $x_{\rm c}\simeq 0.89$, and $p\simeq
1.86$.  This relation gives the system-size dependence of the
critical depinning force, and the divergence of the traveling
distance in any finite system with an exponent $p$, different from
that in an infinite system.

The successful finite-size scaling plots in Fig.~2 including the
exponent $\nu$ imply the divergent correlation length both above and
below the critical depinning force with the same exponent. In
contrast, absence of diverging correlation length below the critical
depinning force was discussed for an elastic string embedded in a 2D
random potential landscape \cite{Giamarchi}. The reason for this
discrepancy remains as an interesting problem for future work.

\vspace{3mm}

\begin{figure}[t]
\psfig{figure=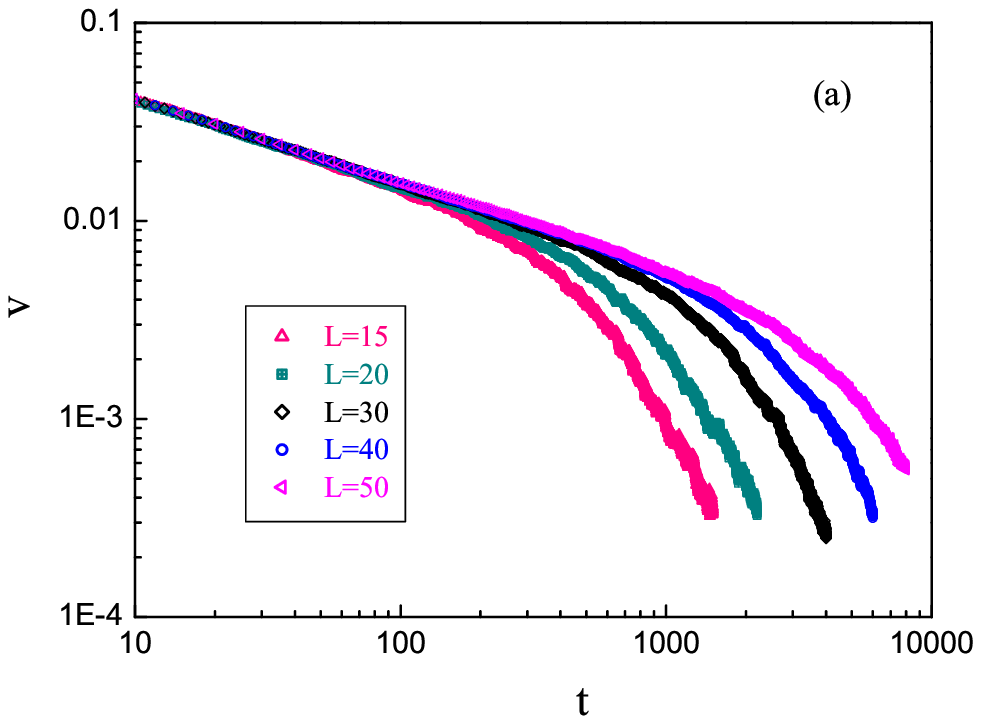,width=8cm}
\psfig{figure=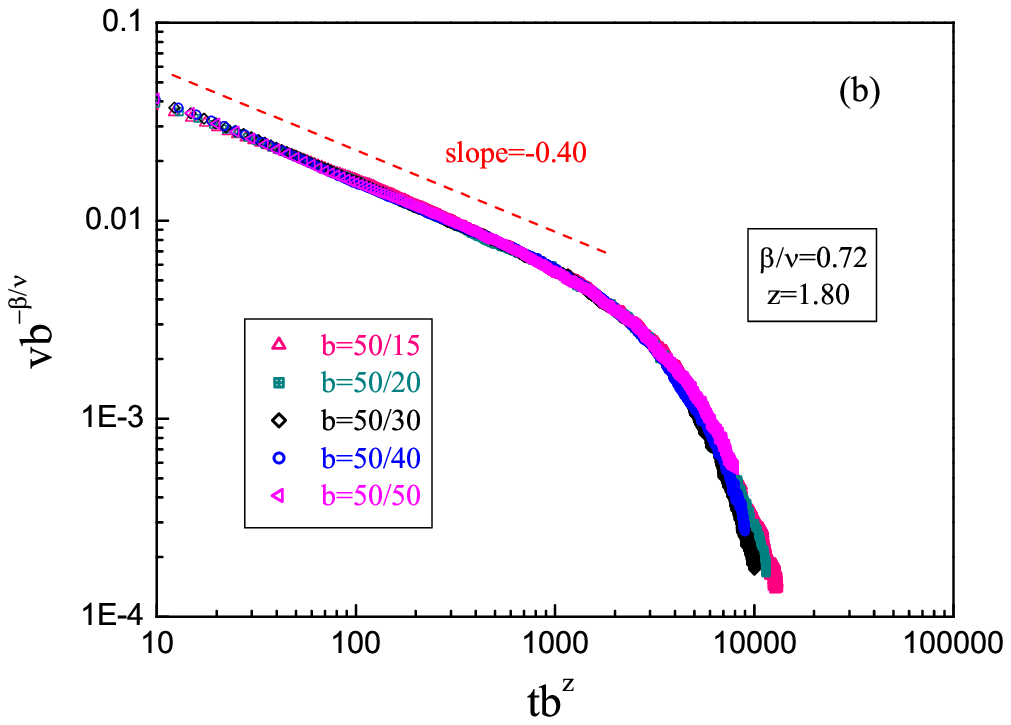,width=8cm} \caption{(color online). (a)
Critical slowing down simulated in several finite systems. (b)
Finite-size scaling plot according to Eq.(\ref{vscaling}). Numbers
of samples are 900-1200 depending the system size.}
\end{figure}

\noindent \textit{Critical slow down. --} As a second transient
process associated with dc driving, we study the critical slowing
down of flux lines, which provides a way to estimate separately the
exponent $z$. Here a steady state of flux lines under driving force
$F=0.6\gg F_{\rm c0}$ is prepared. At $t=0$  the driving force is
reduced to $F_{\rm c0}$. The velocity of the flux lines then
decreases with time according to $v\sim t^{-\beta/\nu z}$
\cite{Janssen}. We have simulated the critical slowing down for
several finite systems, and the results are depicted in Fig.~3(a).
The velocity decreases quickly in a small system since the
corresponding critical force is large, as seen in the onset process
of collective pinning in Fig.~2(a). The finite-size scaling
behavior at $F=F_{\rm c0}$ is described by \cite{Janssen,BZheng}

\vspace{-3mm}
\begin{equation}
v(t,L)=b^{\beta/\nu}v(b^zt,bL).
\label{vscaling}
\end{equation}

\noindent The scaling plot is performed successfully as shown in
Fig.~3(b), and we obtain $\beta/\nu=0.72\pm 0.02$ and $z=1.80\pm
0.03$. The slope of the scaling function at small argument is given
by $\beta/\nu z\simeq 0.40$ as required by the scaling theory
on short-time dynamics \cite{Janssen,BZheng}.

\vspace{3mm}

\noindent \textit{Discussions.--}  The critical exponents estimated
in the present study satisfy the scaling law
 $\beta+\nu(2-z)=1$ \cite{LHTang}.  The roughness
 exponent $\zeta$ can be evaluated
by the scaling relation $\nu=1/(2-\zeta)$ \cite{LHTang} as
$\zeta=1.06\pm 0.04$.

In conclusion, with the aid of computer simulation and
scaling theory, a comprehensive picture has been obtained for the depinning
transition of current driven flux lines in type II superconductors with random
point-like pins.

\vspace{3mm}

\noindent \textit{Acknowledgements.--} We thank T.~Nattermann and S.~-Z.~Lin
for fruitful discussions. This work is supported by WPI Initiative
on Materials Nanoarchitectonics, MEXT of Japan, and Grants-in-Aid
for Scientific Research (No.22540377), JSPS, and partially by CREST,
JST.

\end{document}